\documentclass[aps,prl,final,twocolumn,letterpaper]{revtex4}

\usepackage{graphicx}   
\usepackage{import}                         
\usepackage{epstopdf}
\usepackage{amsmath} 
\usepackage{float} 
\usepackage{bm}
\usepackage{amssymb}
\usepackage{quotes}
\usepackage{indentfirst}
\usepackage{color}
\usepackage{transparent}
\usepackage{dcolumn}
\usepackage{braket}
\usepackage{multirow}
\usepackage{cancel} 
\usepackage{mdframed}
\usepackage{color}
\usepackage{bm}
\usepackage{dsfont}
\usepackage{slashed}
\usepackage{soul, color} 
\soulregister\ref{7}  
\soulregister\cite{7} 
\renewcommand{\st}[1]{}

\usepackage{xr}
\usepackage{amssymb}
\usepackage{amsmath}
\usepackage{mathtools}

\makeatletter
\newcommand*{\addFileDependency}[1]{
  \typeout{(#1)}
  \@addtofilelist{#1}
  \IfFileExists{#1}{}{\typeout{No file #1.}}
}
\makeatother

\usepackage{textcomp} 
\usepackage{xifthen}
\usepackage{xcolor}
\usepackage{etoolbox}
\newboolean{togglechanges} 

\setboolean{togglechanges}{false}

\newcommand{\comment}[1]{\ifbool{togglechanges}
    {#1}  
    {\textcolor{blue}{#1}}}

\usepackage{bibentry}

\usepackage{graphicx}
\usepackage{dcolumn}
\usepackage{bm}

\usepackage{textcomp} 

\begin{document}
\rmfamily

\title{Computational metaoptics for imaging}

\author{Charles~Roques-Carmes$^{1}$}
\email{chrc@stanford.edu}

\author{Kai~Wang$^{2}$}
\author{Yuanmu~Yang$^{3}$}

\author{Arka Majumdar$^{4,5}$}

\author{Zin~Lin$^{6}$}
\email{zinlin@vt.edu}

\affiliation{$^{1}$ E. L. Ginzton Laboratory, Stanford University, 348 Via Pueblo, Stanford, CA 94305\looseness=-1}

\affiliation{$^{2}$ Department of Physics, McGill University, 3600 rue University, Montreal, Quebec H3A 2T8, Canada\looseness=-1}
\affiliation{$^{3}$ State Key Laboratory of Precision Measurement Technology and Instruments, Department of Precision Instrument, Tsinghua University, Beijing 100084, China\looseness=-1}
\affiliation{$^{4}$ Department of Electrical and Computer Engineering, University of Washington, Seattle, Washington 98195, United States
\looseness=-1}
\affiliation{$^{5}$ Department of Physics, University of Washington, Seattle, Washington 98195, United States
\looseness=-1}
\affiliation{$^{6}$ Bradley Department of Electrical and Computer Engineering, Virginia Tech, Blacksburg, VA, USA\looseness=-1}



\clearpage

\renewcommand{\sp}{\sigma_+}
\newcommand{\pbit}{$p$-bit}
\newcommand{\pbits}{$p$-bits}
\newcommand{\sm}{\sigma_-}

\vspace*{-2em}



\begin{abstract}
Metasurfaces—ultrathin structures composed of subwavelength optical elements—have revolutionized light manipulation by enabling precise control over electromagnetic waves' amplitude, phase, polarization, and spectral properties. Concurrently, computational imaging leverages algorithms to reconstruct images from optically processed signals, overcoming limitations of traditional imaging systems. This review explores the synergistic integration of metaoptics and computational imaging, ``computational metaoptics,'' which combines the physical wavefront shaping ability of metasurfaces with advanced computational algorithms to enhance imaging performance beyond conventional limits. We discuss how computational metaoptics addresses the inherent limitations of single-layer metasurfaces in achieving multifunctionality without compromising efficiency. By treating metasurfaces as physical preconditioners and co-designing them with reconstruction algorithms through end-to-end (inverse) design, it is possible to jointly optimize the optical hardware and computational software. This holistic approach allows for the automatic discovery of optimal metasurface designs and reconstruction methods that significantly improve imaging capabilities. Advanced applications enabled by computational metaoptics are highlighted, including phase imaging and quantum state measurement, which benefit from the metasurfaces' ability to manipulate complex light fields and the computational algorithms' capacity to reconstruct high-dimensional information. We also examine performance evaluation challenges, emphasizing the need for new metrics that account for the combined optical and computational nature of these systems. Finally, we identify new frontiers in computational metaoptics which point toward a future where computational metaoptics may play a central role in advancing imaging science and technology.
\end{abstract}

\maketitle

\section{Introduction} 

All imaging systems rely on a combination of optical hardware -- that routes photons from a scene onto a detector -- and software -- that processes the measured signal to generate an image. Both components have been the topic of constant innovation over the past few decades.  

\begin{figure*}
    \centering
    \includegraphics[width=1\linewidth]{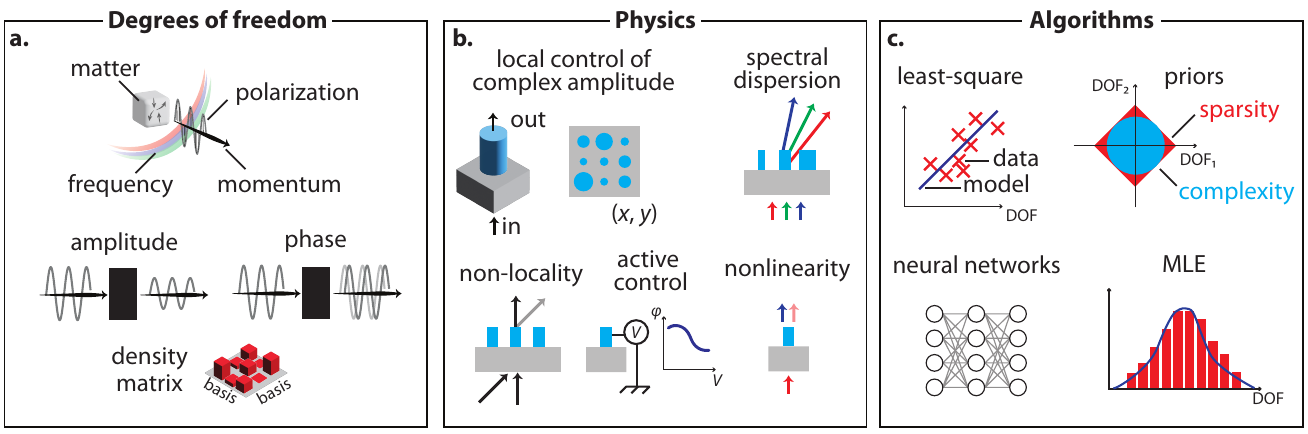}
    \caption{\textbf{Computational imaging with metaoptics: degrees of freedom, physics, and algorithms.} \textbf{a.} The general goal of a computational imaging device is to reconstruct various degrees of freedom of an incident light field, for instance its polarization, frequency, momentum, and complex amplitude distribution. Advanced degrees of freedom (e.g., density matrix of the quantum state of light) may also be of interest. \textbf{b.} Light manipulation is realized by leveraging physical properties of metaoptical devices, such as their ability to locally control the complex amplitude of an incoming wavefront, engineered spectral dispersion, non-locality (spatial dispersion), active control of physical properties (e.g., complex transmission), and nonlinear optical properties. \textbf{c.} Once imaged by a detector, the signal may be reconstructed using various reconstruction and estimation methods, such as least-square error minimization (which may include priors on the reconstructed degrees of freedom, such as high sparsity or low complexity). Black-box methods, such as fully connected neural networks, may also be utilized to classify detected signals and images. Other parameter estimation methods, such as maximum likelihood estimation (MLE) may also be used to estimate the degrees of freedom of the incident light field.}
    \label{fig:overview}
\end{figure*}

On the \textit{hardware} side, metasurfaces -- subwavelength arrays of nanostructured optical elements -- have revolutionized the field of nanophotonics in the past decade~\cite{kuznetsov2024roadmap, yu2014flat, khorasaninejad2016metalenses, yu2011light, genevet2017recent, khorasaninejad2017metalenses, arbabi2015dielectric}. Metasurfaces allow the control of virtually all properties of an incident electromagnetic wave at the nanoscale, such as its polarization, spectral, and angular distribution (Fig.~\ref{fig:overview}b). The development of integrated nonlinear material platforms has also enabled frequency conversion and quantum optical state generation with metasurfaces~\cite{li2017nonlinear, Wang2022metasurfaces, solntsev2021metasurfaces}.

On the \textit{software} side, computational imaging is an interdisciplinary field that combines elements of computer science, optics, signal processing, and imaging technologies to enhance the quality and capabilities of imaging systems. By using algorithms to process and interpret data captured by sensors, computational imaging extends beyond the limitations of optics-only approaches, enabling applications such as high-resolution imaging, volumetric, depth imaging, and advanced object recognition. A foundational aspect of computational imaging is the use of computational methods to reconstruct images from data that may be incomplete, noisy, or otherwise imperfect (some examples shown in Fig.~\ref{fig:overview}c). This approach can involve numerical methods in machine learning, inverse problems, and optimization~\cite{bertero2021introduction, donoho2006compressed, barbastathis2019use}.  

Recent works have explored the fruitful intersection between metaoptics and computational imaging~\cite{lin2021end, lin2022end, huang2022fullcolor, saragadam2024foveated, whitehead2022fast, colburn2018metasurface, li2024singleshot}. This Perspective aims at highlighting this novel and exciting direction for metasurface research which also represents a paradigm shifting opportunity for imaging and sensing technologies. But why is there a natural ``marriage'' between metaoptics and computational imaging? We focus on four key arguments to answer this question below: 

(1) the performance of multifunctional metasurfaces will inevitably reach a ceiling. While single-layer metasurfaces are now able to process multiple spectral, polarization, and momentum degrees of freedom, integrating multiple functionalities in a single device is usually done at the price of key performance metrics (e.g., focusing efficiency, Strehl ratio, etc.). An alternative solution is to stack multiple layers of metasurfaces to lift the limitations of single layers while conserving a compact form factor. For instance, realizing the same amount of functionality as a conventional plan achromatic objective requires a dozen such layers~\cite{lin2021computational}. Proof-of-concept experiments with few layers have been realized at telecom wavelengths~\cite{zhou2018multilayer, roques2022toward}, and with volumetric designs at longer wavelengths~\cite{camayd2020multifunctional, roberts20233d, ballew2023multi}, but realizing nanoscale volumetric patterning at optical frequencies remains an incredible technical challenge; 

(2) many iterative algorithms already used in computational imaging rely on numerical preconditioners that transform the input to facilitate the task of a reconstruction problem~\cite{saad2003iterative}. Metasurfaces are physical preconditioners that can implement simple computational imaging tasks in the optical domain~\cite{long2021isotropic, abdollahramezani2020meta, pors2015analog, cordaro2023solving, koenderink2015nanophotonics, kwon2018nonlocal}. In other words, metasurfaces preprocess the light field from a scene incident on a  detector, whose signal is input to a numerical image-reconstruction task. Optimization of metasurface designs even allows the automatic discovery of optimal preconditioners that do not require human specification. 


Moreover, we argue that synergetic endeavors between metaoptics and computational imaging are natural and effortless: 

(3) any practical imaging application relies on a digital detector whose signal is processed by a computing unit. The use of software for denoising, signal processing, and segmentation is already pervasive, and significant effort has been invested in making application-specific computing units (such as graphics processing units, field programmable gate arrays, and image signal processors~\cite{owens2007survey, bailey2023design, nakamura2017image}) to match the bandwidth and energy requirements of image processing tasks; 

(4) while optimization is already prevalent in image processing, automatic discovery of optimal metasurface designs (with methods such as inverse design, topology optimization, and surrogate models~\cite{jensen2011topology, molesky2018inverse, pestourie2018inverse}) has permeated the field of metaoptics. Computational metaoptic imaging can therefore rely on the natural co-integration of optimization methods from computational imaging and metaoptics design.

This Perspective aims at providing a comprehensive exploration of computational imaging empowered by metaoptics. We begin by revisiting the foundational principles of computational imaging with metasurfaces, underscoring their unique advantages over traditional lens-based systems. We then delve into the end-to-end (inverse) design of computational metaoptics, highlighting how the seamless co-design of hardware and software through gradient backpropagation can lead to superior imaging performance. Our discussion extends to advanced applications enabled by this approach, including phase imaging and quantum photonic state measurement. We also address the challenges of performance evaluation specific to computational metaoptics, proposing suitable metrics for assessing these systems that combine hardware and software. Finally, we outline new frontiers in the field, pointing toward future research directions and opportunities in this rapidly evolving domain.

\section{Revisiting computational imaging with metasurfaces}
Compared to traditional lens-based imaging systems, computational imaging systems expand the realm of imaging hardware possibilities. With the aid of image reconstruction algorithms, they can now form images through media that were previously challenging, such as optical fibers \cite{amitonova2018compressive} or diffuse white walls \cite{faccio2020non}. Moreover, computational imaging captures not just the two-dimensional (2D) intensity of a scene but also additional information, including three-dimensional (3D) profiles, polarization, and spectral signatures.

Given the variety of existing computational imaging systems, one might wonder what unique advantages their integration with metaoptics technology could offer. Here we highlight two key benefits:

\textit{Subwavelength, multifunctional optical control} : Metasurfaces provide finely controlled, multifunctional optical responses that are easily calibrated. Unlike conventional diffractive optical elements, metasurfaces manipulate the wavefront at a sub-wavelength scale, which is ideal for large-angle light illumination, beam bending, and focusing \cite{ni2020metasurface}. Each meta-atom can be tailored with a wide range of polarization and spectral responses \cite{wen2024metasurface}. By interweaving different meta-atoms \cite{maguid2017multifunctional}, multifunctional responses can be achieved. Metasurfaces also offer more precise control and easier calibration than other media like optical fibers or random diffusers. Their optical responses are meticulously designed using full-wave electromagnetic simulations, with fabrication precision at the nanometer level. This contrasts with the more holistic, less flexible control of optical fibers or random diffusers, which can complicate system calibration.

\textit{Scalable fabrication} : While metasurfaces were once fabricated using costly electron-beam lithography, recent advancements have made scalable, wafer-level manufacturing possible~\cite{kuznetsov2024roadmap}. Techniques like nano-imprinting and deep-UV lithography are now being used by major technology companies and startups worldwide. This progress moves metasurface technology from theoretical to practical, with a clear path to leveraging established semiconductor manufacturing processes for cost-effective, large-scale production.

Leveraging these capabilities, a common strategy has been to design metasurface hardware and image reconstruction algorithms independently. A simple example is that of a metalens placed at the aperture plane of an imaging system, which can be designed for specific tasks (parametrized by a transfer function). To enable single-particle tracking, the pupil function encoded by the metasurface can be tailored to create a point-spread function (PSF) sensitive to the distance of a point light source \cite{jin2019dielectric,colburn2020metasurface}, with accuracy gauged by Fisher information or the Cramér-Rao  bound (see Section on performance metrics). 

Similarly, metalenses can be directly designed for polarization \cite{arbabi2018full,zhao2021metalens, rubin2019matrix}, spectral \cite{froch2022dual}, or multi-dimensional \cite{shen2023monocular,wang2024metasurface} imaging. 
Alternatively, placing a metasurface filter array at the sensor plane allows for the creation of a measurement matrix optimized for spectral imaging \cite{wang2009angle,xiong2022dynamic}. Image reconstruction can then be performed using conventional iterative optimization or deep learning techniques. 
While these design methods leverage the multifunctional capabilities of metasurface imaging to realize computational imaging, recent works -- which we describe in the next section -- have instead adopted an ``end-to-end'' approach that designs metasurfaces and the computational reconstruction algorithms \textit{simultaneously}.

\section{End-to-end (inverse) design of computational metaoptics}
\begin{figure*}
    \centering    \includegraphics[width=1\linewidth]{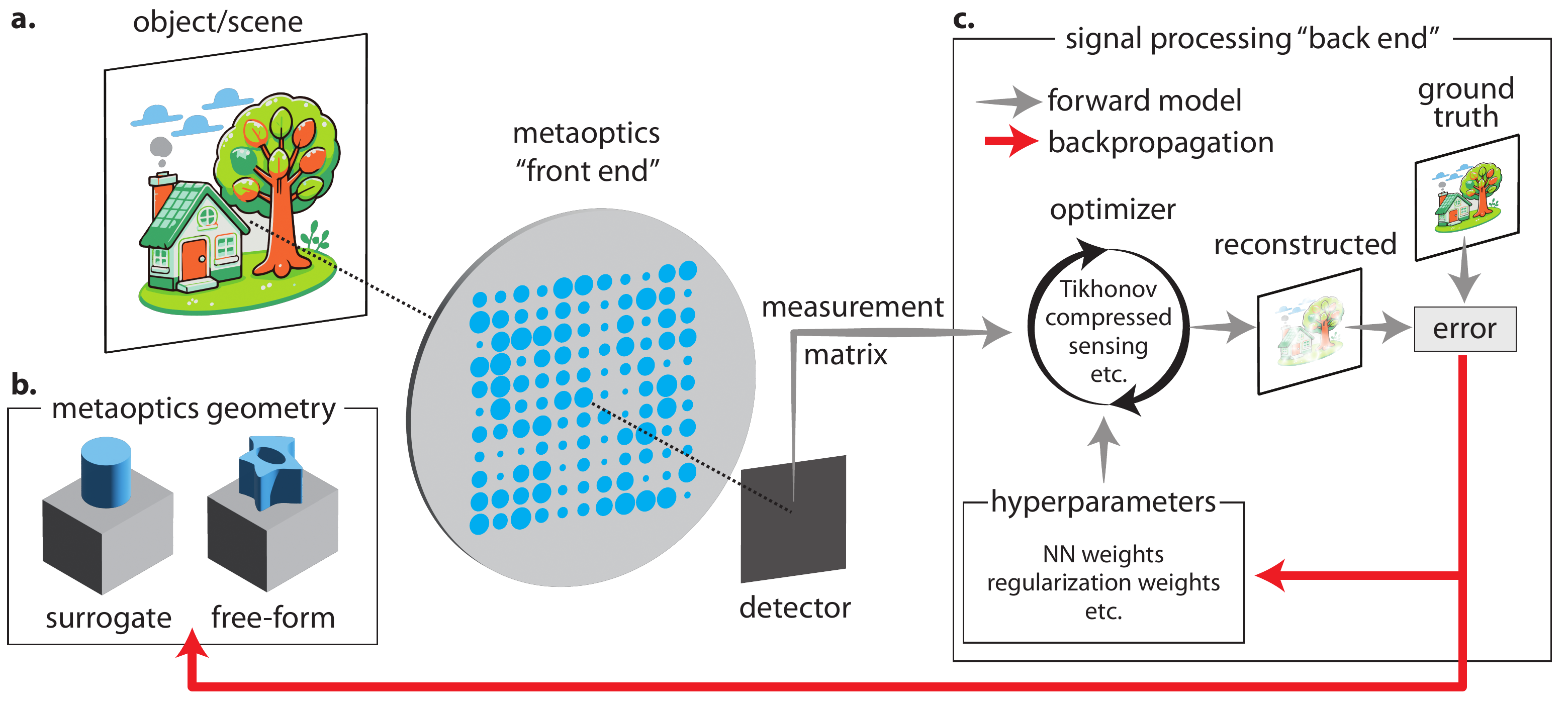}
    \caption{\textbf{End-to-end computational metaoptic imaging via synergistic inverse design of geometrical degrees of freedom and hyperparameters.} \textbf{a.} Light emanating from a scene or object is processed with a metaoptics (corresponding to the system's ``front end'') and imaged onto a detector. Additional optical or electronic components (not shown) may be used to further process light or the generated signal on the detector. \textbf{b.} The metaoptical device is defined by its geometry, which can be optimized via surrogate models or free-form (topology) optimization, or a combination thereof. \textbf{c.} The generated signal is processed by an optimizer (or estimator, corresponding to the system's ``back end''), generating a reconstructed object. The calculated error signal is back-propagated to adjust reconstruction hyperparameters and metaoptics geometry, in order to reduce the error.}
    \label{fig:end2end}
\end{figure*}

Instead of independently developing the optics and the reconstruction algorithm, one may co-design hardware and software seamlessly by end-to-end gradient backpropagation. The general framework for end-to-end inverse design is shown in Fig.~\ref{fig:end2end}. This design method relies on the simultaneous optimization of physical nanophotonic degrees of freedom (parametrized by $p$) and image reconstruction hyperparameters $\alpha, \beta$ (e.g., corresponding to weights of a neural network, or regularization parameters of iterative reconstruction algorithms). 

Generally, end-to-end design of metaoptics computational imaging can be formulated using a bi-level optimization problem~\cite{lin2022end} (Fig.~\ref{fig:end2end}), defined by the following set of equations:
\begin{widetext}
\begin{align}
    \text{(End-to-end objective function)} \quad  \quad  & \min_{p,\alpha,\beta} ~ L (p,\alpha,\beta) =
    \langle \lVert u_0 - u_\text{est} (p,\alpha,\beta)\rVert^2 \rangle_{u_0,\eta}, \\
    \text{(Nested image reconstruction)} \quad  \quad & u_\text{est} = \arg \min_u \lVert G(p)u-v \rVert^2 + \sum \alpha_i \lVert A_i u \rVert^2 + \sum \beta_i |B_iu|_1, \label{eq:nested} \\
    \text{(Image formation)} \quad  \quad &  v = G(p)u_0 + \eta, \quad G\sim |E(r_\text{sensor},\lambda)|^2, \label{eq:nested3}\\
    \text{(Maxwell physics)} \quad  \quad &  \nabla \times \nabla \times E(r)-\omega^2 \varepsilon(r;p) E(r) = i\omega \delta(r-r_0)\hat{e}, \quad
    \omega ={2 \pi \over \lambda}, ~ 
    \hat{e} \in \{\hat{x},\hat{y},\hat{z}\}. \label{eq:nested4}
\end{align}
\end{widetext}
The primary objective of the above set of equations is to minimize the reconstruction error $L(p,\alpha,\beta)$. Importantly, this function depends simultaneously on the nanophotonic parameters $p$ (such as the geometrical parameters of nanostructures in a large-area aperiodic metasurface) \textit{and} the image reconstruction hyperparameter(s) $\alpha,\beta$. The cost function $L$ is averaged over an ensemble of training objects (with ground truth $u_0$) and noise realizations $\eta$. For each $u_0$, the estimate $u_\text{est}$ is obtained by solving a nested minimization problem (Eq.~(\ref{eq:nested})) involving a data-fitting term $\lVert G u-v \rVert^2$ and additional regularization terms with weights $\alpha, \beta$. In Eq.~(\ref{eq:nested3}), $v=G u_0 +\eta$ represents the noise-corrupted raw image, and the measurement matrix $G(p)$ embodies the physics of image formation (parametrized by nanophotonic parameters $p$). 

Accurate computations of $G$ and the gradient $\nabla_\varepsilon G$ require detailed simulations of the nanophotonic permittivity $\varepsilon(r;p)$ using fullwave vectorial Maxwell equations and their adjoint~\cite{molesky2018inverse}, taking into account strong spatial, spectral and polarization dispersions (see Eq.~(\ref{eq:nested4})). In fact, exploiting the richness and complexities of $G$ that arise from the Maxwell equations categorically differentiates end-to-end metaoptics from diffractive and refractive optics as well as from popular data-driven approaches~\cite{sitzmann2018end,baek2020end,yang2024end,fu2024limitations}. It is also important to emphasize that Eq.~(\ref{eq:nested}) represents a regularized regression problem, which is itself embedded within the primal problem of minimizing $L$. The regularized regression enforces explicit conditions on $u_\text{est}$, such as uniqueness, stability, smoothness, non-negativity, and/or sparsity, and can be handled by conventional iterative algorithms with provable convergence and correctness. During the back-propagation stage, the derivative of $u_\text{est}$ is efficiently computed by an adjoint sensitivity analysis of the Karush-Kuhn-Tucker optimality conditions~\cite{lin2021computational}. While conventional regression back end (Eq.~(\ref{eq:nested})) is fully interpretable and works well for explicitly defined priors, it may be replaced or augmented by an artificial neural network to learn implicit priors from training data. However, care must be taken to ensure generalizability and resilience against noise and adversarial attacks. It was recently shown that hybrid architectures~\cite{tseng2021neural} could outperform pure deep learning back ends.
\begin{figure}
    \centering    
    \includegraphics[width=0.9\linewidth]{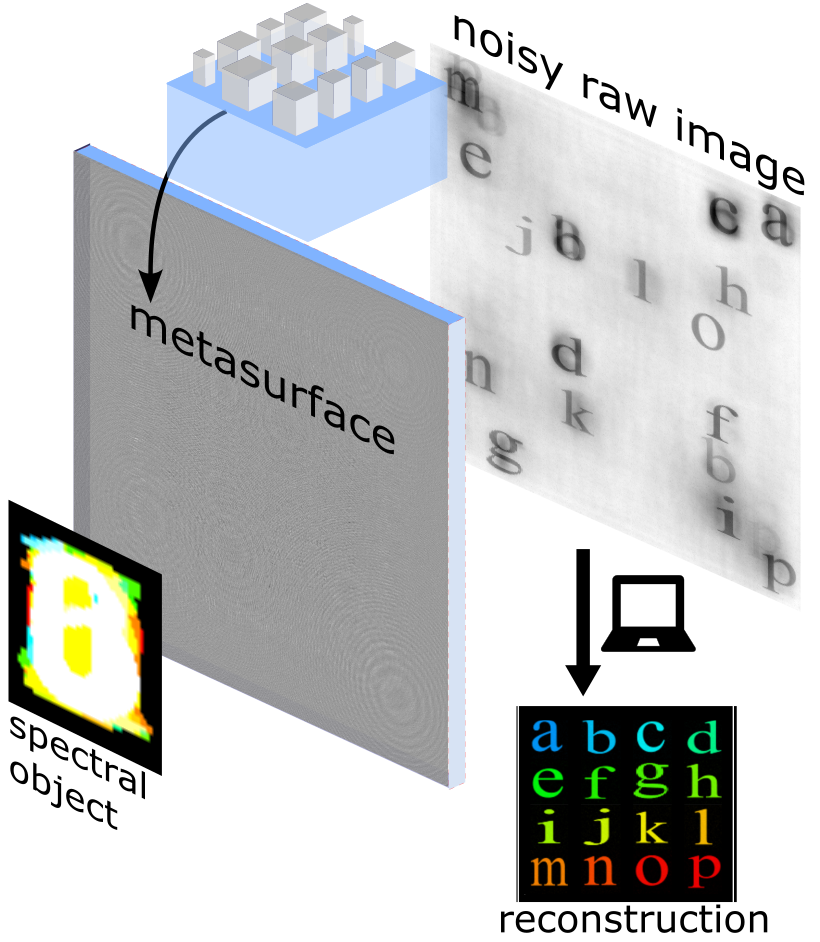}
    \caption{\textbf{End-to-end optimized multi-spectral imaging.} A metasurface captures a multispectral object (left) and produces a greyscale image (500x500 pixels), from which  a computer reconstructs the multispectral ground truth (16 spectral channels, each with 50x50 pixels).}
    \label{fig:e2e_specimg}
\end{figure}

In contrast to optics-only or computation-only designs, the operating principle of the front end optics is not prescribed \textit{a priori} but is allowed to emerge spontaneously from the end-to-end optimization described by Eqs.~(1-4). For example, it was recently shown that an end-to-end optimization of a single-layer metasurface in conjunction with a simple $\ell_2$-regularized least-square regression could enable filter-free snapshot multi-spectral imaging~\cite{lin2022end}. Fig.~\ref{fig:e2e_specimg} illustrates an example application of this end-to-end design method for multi-spectral imaging: a $0.6\times 0.6$ mm$^2$ metasurface made up of two million TiO$_2$ nano-pillars on a silica substrate optimized for 16-color spectral imaging. The optimized metasurface successfully images superimposed letters, each emitting a different wavelength, which cannot be distinguished by the naked eye or a conventional imaging lens. The resulting single-shot monochrome image showcases the spatial demultiplexing capability of the end-to-end optimized metasurface, effectively separating and focusing light across different wavelengths. The positions of the foci, which emerge spontaneously from the end-to-end optimization, are highly non-intuitive as opposed to a human-designated regular imaging pattern, e.g., a lattice of focal spots. These \textit{emergent} foci greatly improve the noise tolerance of the regression problem while maximally exploiting the metasurface’s intrinsic dispersions and sensitivities. As such, these focal positions cannot be pre-designated by a metaoptics-only approach. Notably, unlike deep learning methods, the bi-level optimization approach is essentially \emph{data-agnostic}: it requires minimal training data (less than 30 ``random dots'' ground truths) yet generalizes perfectly to any scene thanks to the fully interpretable imaging mechanism from Eqs.~(1-4), which spontaneously emerges from the end-to-end optimization.

\section{Advanced applications enabled by computational metaoptic imaging}

\begin{figure}
    \centering
    \includegraphics[width=1.9\linewidth]{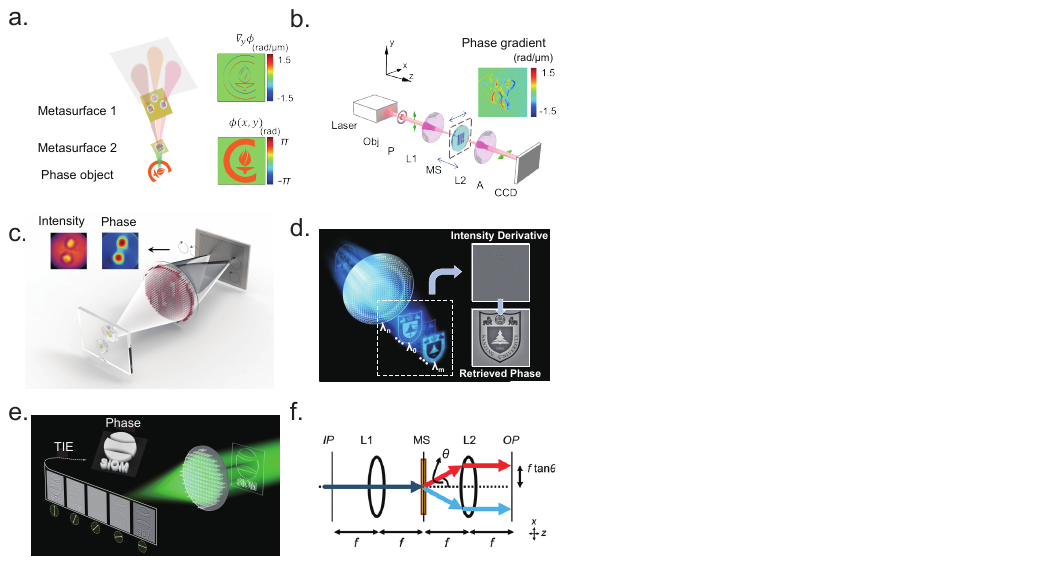}
    \caption{\textbf{Advanced applications enabled by computational metaoptic imaging.} \textbf{a.} A compact quantitative phase gradient microscope based on 3-step phase shifting using a spatially-multiplexed bi-layer metasurface. Reproduced with permission from Ref.~\cite{kwon2020single}. \textbf{b.} A Fourier optical spin splitting microscope by inserting a polarization-multiplexed metasurface into the Fourier plane of a conventional 4$f$ imaging system. Reproduced with permission from Ref.~\cite{zhou2022fourier}. \textbf{c.} Single-shot deterministic complex amplitude imaging with a single-layer metalens based on polarization phase-shifted shearing. Reproduced with permission from Ref.~\cite{li2024single}. \textbf{d.} A mechanical scanning-free transport-of-intensity-based quantitative phase imaging microscope using a spectrally-dispersive metalens. Reproduced with permission from Ref.~\cite{wang2024quantitative}. \textbf{e.} A mechanical scanning-free transport-of-intensity-based quantitative phase imaging microscope using a polarization-multiplexed metalens. Reproduced with permission from Ref.~\cite{min2024varifocal}. \textbf{f.} A single-shot transport-of-intensity-based quantitative phase imaging microscope using a polarization-multiplexed metasurface. Reproduced with permission from Ref.~\cite{engay2021polarization}.}
    \label{fig:app}
\end{figure}

We now move on to advanced applications of metaoptic computational imaging driven by combined innovations in metaoptic design and physics-aware reconstruction algorithms.

\subsection{Phase imaging}

Traditional imaging systems are limited to capturing light intensity, yet the phase information, which can reveal profound insights into an object's physical and chemical properties, is crucial (see degrees of freedom that describe incident wavefronts in Fig.~\ref{fig:overview}). This has led to widespread applications in fields such as optical metrology \cite{zuo2022deep}, adaptive optics \cite{guyon2005limits}, and biomedical imaging \cite{park2018quantitative}.

Interferometry has been the go-to method for phase detection, but its bulky setups and sensitivity to vibrations have been a challenge. Over the past decade, innovative computational phase retrieval techniques—such as transport-of-intensity (TIE)~\cite{zuo2020transport} and Fourier ptychography~\cite{zheng2013wide} -- have emerged, significantly streamlining hardware requirements. However, these methods face their own hurdles, including the need for multiple measurement frames and reliance on specific sample priors.

Recently, it has been shown that one may address these issues with existing phase imaging systems using metaoptics -- leveraging the rationale that metaoptic can impart complex optical functions in a compact form factor. For instance, one can simplify conventional interferometric setups by using a spatially-multiplexed bi-layer metasurface. This metasurface realizes a compact 3-step phase shifting setup, and can be used for quantitative phase gradient microscopy~\cite{kwon2020single} (Fig.~\ref{fig:app}a). Later, 4-step phase-shifting interferometric setups have also been realized by simply inserting a polarization-multiplexed metasurface into the Fourier plane of a conventional 4$f$ imaging system \cite{zhou2022fourier,liu2023computing} (Fig.~\ref{fig:app}b). However, in the abovementioned cases, with a single measurement, one can only obtain one-dimensional (1D) phase gradient information, which leads to ambiguity in reconstructing the 2D phase distribution. This challenge can be overcome by combining a single-layer polarization- and spatially-multiplexed metalens with a polarization camera to simultaneously record four polarization shearing interference patterns along both in-plane directions, thus allowing the deterministic reconstruction of the complex amplitude light field in a single shot~\cite{li2024single} (Fig.~\ref{fig:app}c).

Another common method in phase microscopy is TIE, which can also be enhanced using metaoptics. The TIE technique, widely studied for computational phase retrieval, typically requires multiple measurements through axial scanning of the target object. A TIE-based phase imaging system that eliminates mechanical movement was realized, by using a spectrally-dispersive metalens to capture multi-focal images via varying illumination wavelengths \cite{wang2024quantitative} (Fig.~\ref{fig:app}d). This concept was then further developed into a varifocal metalens-based phase imaging system, achieving focal length scanning by altering the polarization state of the incident light \cite{min2024varifocal} (Fig.~\ref{fig:app}e). Another approach consists in utilizing a polarization-dependent metasurface to capture two images with different focal lengths, constructing a single-shot TIE-based phase imaging system \cite{engay2021polarization} (Fig.~\ref{fig:app}f).

Metaoptics has the potential to greatly simplify existing phase imaging systems -- which typically rely on bulky free-space interferometric setups -- while maintaining robust phase retrieval capabilities. Given the important applications of phase imaging in biology, we envision that this could pave the way for unique applications in point-of-care diagnostics for fast-moving targets and in space-constraint environments like endoscopy~\cite{shanker_quantitative_2024}.

\subsection{Quantum photonic state measurement}


The direct access to all the degrees of freedom of light with metaoptics also provides a unique and powerful means to measure quantum states of light~\cite{Wang2022metasurfaces} (see Fig.~\ref{fig:overview}). In this subfield of computational metaoptics, the quantum optical states of interest are, e.g., entangled photons represented in polarization, spatial, and spectral degrees of freedom. The metaoptics plays the role of an unconventional quantum ``camera lens'' that performs a linear transformation to the incoming beam of photons. The image sensor in classical computational imaging is then replaced by arrays of single-photon detectors or single-photon-sensitive image sensors. 

\begin{figure}
    \centering
    \includegraphics[width=1\linewidth]{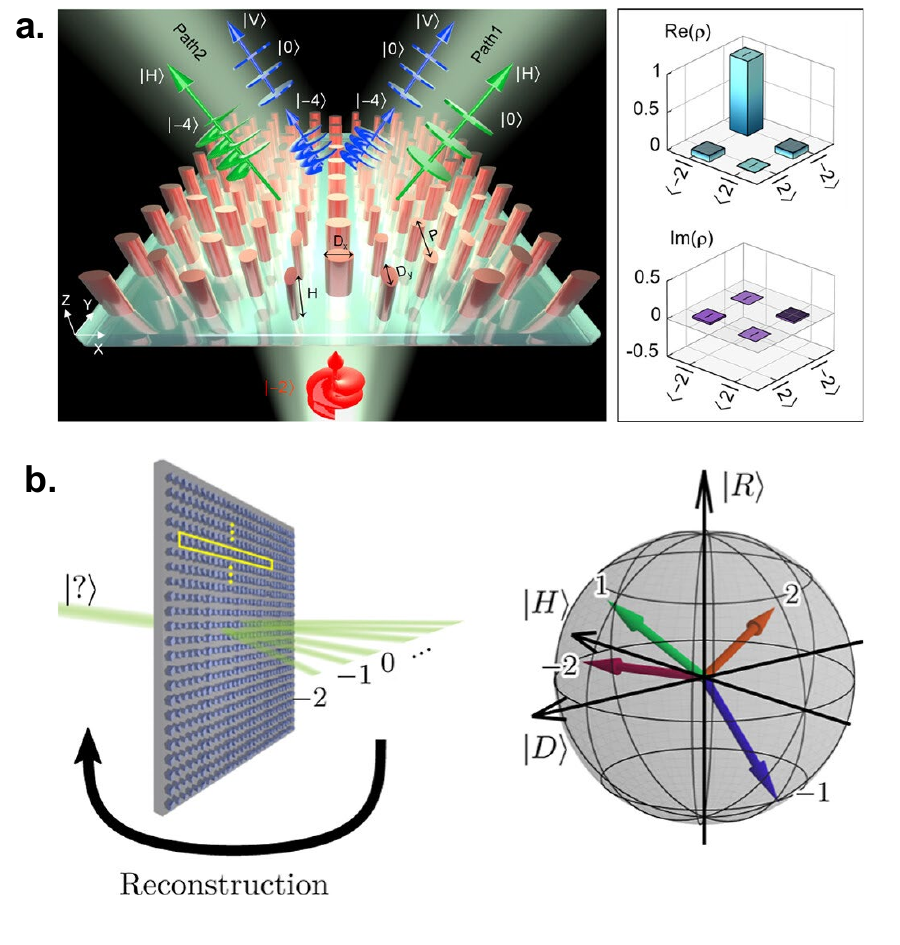}
    \caption{\textbf{Two recent examples of metaoptics for quantum photonic state measurement and reconstruction.} 
    \textbf{a.} A metasurface for reconstructing density matrices represented in the orbital angular momenta (OAM) basis. Reproduced with permission from Ref.~\cite{Wang2023characterization}. \textbf{b.} A metasurface composed of a single periodic meta-grating that can perform an optimized set of generalized polarization measurements represented by general POVMs, which can robustly reconstruct polarization density matrices. Reproduced with permission from Ref.~\cite{Lung2024robust}. 
    }
    \label{fig:quantum}
\end{figure}

Quantum state tomography -- measurement of the density matrix of an incoming quantum state -- typically relies on a set of projective measurements onto different subspaces (yielding reduced density matrices), followed by computational reconstruction of the full density matrix. It is therefore a measurement method that naturally lends itself to the framework of computational metaoptics. Some of the recent schemes for quantum tomography with metaoptics are shown in Fig.~\ref{fig:quantum}.
It was first realized that a dielectric metasurface that can diffract photons based on their polarization state~\cite{wang2018quantum} can be used to form a tomographically complete set of projective measurements of the multiphoton polarization state. The set of projective measurements is optimized in a way that the reconstruction problem is near-optimally conditioned with minimal error amplification in the reconstruction. More specifically, such an optimization is done by minimizing the condition number of the instrument matrix that describes the transformation that relates the the measured quantities (here being the elements in an $N$-photon polarization-state density matrix) to the detections (here being the $N$-photon correlations across the detectors). We provide more details on the use of the measurement matrix's condition number as a metric of performance evaluation in the next section (``Performance evaluation''). 
Such a metasurface-based measurement scheme can replace conventional quantum state tomography methods that requires many reconfigurations of waveplates, hence promises enhanced simplicity, stability, scalability, and miniaturization.

Quantum states of light are often encoded in other degrees of freedom, for example, in spatial modes. A commonly used basis consists of the Laguerre-Gaussian states that correspond to different photon orbital angular momenta (OAM). An advantage of using such a spatial-state basis to encode quantum information lies in the larger Hilbert dimension (compared to the polarization-encoding scheme), making photons ``qudits'' instead of ``qubits.'' Since metaoptics also allows arbitrary linear transformations over spatial mode bases, recent work~\cite{Wang2023characterization} has started to use the quantum state tomography scheme to measure OAM states of photons. We also note that recent work proposed the use of self-configuring optics to realize quantum state tomography of biphoton states in the spectral or spatial domains~\cite{roques2024automated}.

Conventional quantum state tomography is based on a sequence of projective measurements, where each measurement corresponds to a definite (pure) quantum state. For a polarization-encoding scheme, a projective measurement corresponds to a perfect polarizer with an infinite extinction ratio. In practice, the diffracted light from metasurfaces, upon the detection of a polarization-insensitive detector, cannot always be approximated by projective measurements as they are more generally partial polarizers. Such measurements are described as general positive operator-valued measures (POVMs). Recent work~\cite{Lung2024robust} demonstrated a metasurface for polarization state measurement based on a general POVM framework. In this work, a periodic single meta-grating is used, with a certain number of its diffraction orders forming a set of POVMs that are optimized for the reconstruction of quantum and classical polarization states of light. The single meta-grating design minimizes diffraction losses associated with interleaving different meta-gratings for achieving projective measurements~\cite{wang2018quantum}. A general POVM framework has also been used more recently to design metasurfaces performing randomized measurements on polarization states of photons~\cite{Ren2024error}, which can also reconstruct the state in conjunction with error mitigation techniques. Shadow tomography~\cite{aaronson2018shadow, huang2020predicting}, a relatively new approach in quantum state tomography, was also recently brought to metasurface experiments~\cite{An2024efficient}. The shadow tomography approach aims to predict functions of a quantum state instead of reconstructing the full state, with the benefit of using only a logarithmic number of measurements. All of these approaches show that synergetic advances in metaoptics and reconstruction methods can reduce the effective computational complexity of quantum tomography.

\section{Performance evaluation}

Given the expected rise of interest in works at the intersection of metaoptics and computational imaging, and their promise to enhance various fields of imaging, it is important to review performance evaluation metrics. We first review the use of the modulation transfer function (MTF) for characterizing the imaging performance of metaoptics (which is traditionally an ``optics-only'' figure of merit), and then turn our attention to figures of merit that incorporate the performance of reconstruction algorithms. 

\subsection{Modulation transfer function}
One critical aspect of computational imaging is to evaluate the imaging performance of a designed piece of metaoptics. Traditional metrics like diffraction efficiency (e.g., incorporating degradiation due to stray light) are often difficult to measure and calculate for metaoptics, especially for large aperture metaoptics~\cite{banerji2019imaging}. This is tied to the inability to simulate a large-area metaoptics without making significant approximations~\cite{pestourie2018inverse}. For example, the standard stray-light models from ray-optics simulators often underestimate the amount of stray light measured from metaoptics devices. Diffraction efficiency, for traditional diffractive optics is defined as the power going into a specific higher-order diffraction channel~\cite{goodman2015statistical}. In metaoptics with subwavelength unit cells, the lack of higher-order diffraction makes such measurements difficult, and often researchers rely on focusing efficiency. Such focusing efficiency is however not well-defined for non-lens like PSF, which is often the case for computational imaging, and also challenging to measure for metalenses~\cite{menon2023perspectives}. Similarly, the Strehl ratio is not well-defined for computational metaoptics, if the final signal is the outcome of a reconstruction back end. 

An alternative figure of merit for metaoptics imaging is the MTF. A recent work measured the effective MTF of a metaoptical imaging system (with 1~cm aperture) and showed a drastically improved MTF compared to the MTF of the optics itself~\cite{tseng2021neural}.  Based on empirical evidence on broadband imaging, it was found that the key to achieve a high quality imaging is the transmission efficiency of the metaoptics, and an MTF with cutoff (determined by the noise of the sensor) at high spatial frequency~\cite{huang2024broadband}. Said differently, a better imaging performance necessitates high volume under the integrated MTF curve~\cite{froch2024beating}. Depending on applications, we can define a different figure of merit based on MTF. As such, transmission efficiency and MTF are two important metrics for metaoptics co-optimized with a computational back end.

\subsection{Condition number and information measures}
Another useful metric which characterizes the noise sensitivity of the reconstruction process is the condition number $\kappa$ of the imaging matrix $G$ (for linear imaging processes, as seen in Eq.~(\ref{eq:nested3})), defined as the ratio of the largest to the smallest singular values of $G$. 

In general, $\kappa$ provides an upper bound to the potential amplification of image noise by the reconstruction process. However, a direct minimization of $\kappa$ is typically expensive and often intractable; instead, it have shown that end-to-end optimization directly leads to significant reductions of $\kappa$~\cite{lin2021end}, and therefore to greater noise robustness. More sophisticated criteria exist for reconstruction problems with priors and provide, in some instances, theoretical bounds for the imaging reconstruction performance. For example, in the $\ell_1$-regularized compressed sensing problem (which enforces sparsity prior), the theoretically optimal condition for $G$ is the so-called restricted isometry property (RIP)~\cite{donoho2006compressed}. It is well known that RIP is satisfied by Gaussian matrices (i.e., a matrix whose entries are i.i.d. Gaussian distributed), which cannot be physically realized in a snapshot setting. End-to-end optimization has been shown to discover physically realizable imaging matrices $G$, which closely approach the performance of Gaussian matrices in snapshot compressive imaging problems~\cite{arya2024end}. 

For more advanced reconstruction back ends and implicitly learned image priors (such as neural networks), it becomes much more challenging to establish theoretically rigorous criteria for optimal performance. One possible candidate is Fisher Information (FI) and the associated Cramer-Rao's bounds~\cite{ly2017tutorial}, which directly tie the experimental signal-to-noise ratio to the information content of the measuring system. However, FI is intractably expensive to compute for imaging problems and may not be a valid criterion for biased estimators (i.e., with priors). Recently, other information-theoretic approaches, such as mutual information estimates, have shown some promise~\cite{pinkard2024universal} in characterizing computational imaging systems, in conjunction with theoretical bounds~\cite{dong2021fundamental, bouchet2021fundamental}.

The development of an ``ideal'' performance metric would be one that integrates transmission efficiency and MTF with information-theoretic measures like mutual information, as well as explicit ties to experimental performance (such as Cramer-Rao's bounds), providing a comprehensive evaluation of both the metaoptics system and its co-optimized computational reconstruction.

\section{New frontiers in computational metaoptics}

We now discuss several new directions in the field, which we think are unique opportunities enabled by metaoptic computational imaging.

\subsection{Optical encoders}
One of the most important and challenging problems in computer vision is detecting and identifying an object in a scene, which has a profound impact on wearable technologies and autonomous navigation. While the vision of an all-optical pattern recognition has been sought for decades \cite{anderson1986coherent, anderson1986optical}, recent developments in machine learning and metaoptics open up opportunities to revisit this problem. A metaoptical front end can encode information from a scene, which in conjunction with a computational back end can perform object detection~\cite{zheng2022meta, colburn2019optical}. Current efforts in designing metaoptical front ends can be broadly classified into two categories: arbitrary vector-matrix multiplication using multiple optics~\cite{shen2017deep, prabhu2020accelerating, pai2023experimentally, miller2013self, nahmias2019photonic}; PSF engineering to perform pre-defined mathematical operations using a single (meta)optic~\cite{silva2014performing, wang2022design, asgari2024multifunctional, cordaro2023solving}. In both cases, designing the architecture amounts to figuring out how much of the computation will be performed with optics versus how much is handled in the digital domain. 

There are largely two different methodologies that have been employed: in one case, typical deep learning layers are directly mapped onto the optics; and the other approach involves ``end-to-end'' design. In the first approach, researchers have implemented the first layer of a typical deep learning block; however, the performance benefit at a system level remains questionable as benchmarking with existing GPU-based inference shows little benefit in terms of latency and power to implement only the first layer~\cite{colburn2019optical}. One possible avenue is to implement multiple linear layers with optics, by taking the information back and forth between optics (for linear operation) and the digital domain (for nonlinear operation). However, the large latency and power associated with signal transduction (in the display and image sensors) renders such a hybrid approach sub-optimal. Transferring a large number of linear operations to the front end, implementing them with optics, and then performing digital operations could however provide a benefit. 

We emphasize that pure linear operations can also provide some classification in simple systems, e.g., the hand-written dataset in MNIST is linearly separable, and researchers demonstrated classification with only linear operation using multi-layer diffractive optics~\cite{lin2018all}. However, such linear operations are unlikely to provide high classification accuracy in complex datasets and real-life scenes.

To understand the efficacy of an optical front end, a recent work performed end-to-end design of a hybrid metaoptics/digital neural network~\cite{huang2023photonic}. An empirical finding from this work is that an optical front end can be beneficial for computer vision when latency and power are both highly constrained. However, such an advantage is achieved when the overall accuracy is lower than what can be obtained using the best-trained digital neural network. In another work, using an end-to-end designed metaoptical front end \cite{wei2023spatially}, high classification accuracy with the CIFAR-10 dataset was demonstrated. While a purely digital version needs 57M parameters (with a classification accuracy of $\sim$72.64\%), a hybrid optical-digital version requires only 2K parameters while maintaining a similar classification accuracy ($\sim$73.80\%). In that example, the end-to-end design exploits the spatially-varying PSF of the metaoptics to achieve high classification accuracy with a much model size.

\subsection{Computational imaging of light-matter interactions}

The reconstructed signal from an image processing task is usually meant to be human-interpretable, relating to the intensity of scattered or reflected light from an object or scene. Computational imaging and nanophotonics give us access to other fundamental properties of light that are invisible to the human eye (see overview in Fig.~\ref{fig:overview}), such as polarization~\cite{rubin2019matrix}. Since the detected signal is generated by the interaction of oscillating electromagnetic fields with matter (in the object/scene, their interaction with the imaging system, and the detector), metaoptic computational imaging could give us access to fundamental degrees of freedom in imaging : the complete electromagnetic representation of the (quantum) optical fields as well as their interactions with matter~\cite{barbastathis2019use}. 

A direct example of this rationale is in partially coherent light or incoherent light emitters, such as electrically-driven LEDs, scintillators, and thermal emitters~\cite{greffet2018light, roques2022framework}. There, light emission processes originate in fluctuating currents (that relate to electronic occupation levels in the emitting materials). The measured image on the detector can be expressed as:
\begin{equation}
    v = \text{Trace}\left( G_E(p)^\dagger G_E(p) \langle J J^\dagger \rangle \right),
\end{equation}
where $G_E(p)$ is the dyadic Green's function (relating currents to electric fields) imparted by a nanophotonic structure of geometry $p$ (as in Eq.~(\ref{eq:nested3})), and $J$ the stochastic current distribution in the emitter. In the framework of fluctuational electrodynamics, the average current correlations $\langle J J^\dagger \rangle$ relate to electronic occupation levels (e.g., in the valence and conduction band of a semiconductor)~\cite{henry1996quantum}. One class of systems where the measurement of the emitter state is well developed is solid-state qubits (such as single electron and nuclear spin qubits in solid-state color centers~\cite{awschalom2018quantum}), where the quantum state of single qubits can be read out with high fidelity. 
More generally, optimization of the nanophotonic geometry $p$, in conjunction with reconstruction algorithms to extract features of $\langle J J^\dagger \rangle$ -- such as sequential variational optimizers~\cite{roques2024measuring, roques2024automated} -- could yield information about the emission mechanisms (and their spatio-temporal dependence) in these partially coherent light emitters. 

Extending this paradigm, one may also venture into the quantum regime, as we discussed in a previous section of this Perspective. A key promise of computational metaoptics for quantum technologies lies in the measurement of high-dimensional quantum states of photons, e.g., those encoded in many spatial states, frequencies, temporal states, or a combination of different degrees of freedom of photons. While current works remain in a relatively small Hilbert dimension primarily due to the small number of detectors, we anticipate that by combining advanced single-photon image sensors and employing compressed sensing that is capable of reconstructing underdetermined problems with prior knowledge, metaoptics will enable genuine high-dimensional photonic state measurements. On the other hand, the numerical design techniques that have been shown to be powerful in classical computational imaging problems, such as end-to-end optimization, can also further empower metaoptics in quantum state measurements and quantum imaging.

\subsection{Hyperscale end-to-end differentiable models}
It is a long-standing observation that ultra-high dimensional optimization problems (with millions to trillions of optimization parameters) are readily susceptible to gradient-based algorithms. These problems enjoy the so-called blessing of dimensionality: contrary to intuition, having more parameters to optimize---preferably more than ``necessary''---makes the problem easier and not harder~\cite{gershenfeld1999nature,du2019gradient,kawaguchi2016deep,goodfellow2016deep,choromanska2015loss,arora2019fine,lyu2021gradient,trivedi2021gradient}. In recent decades, large-scale optimizations have empowered artificial intelligence and deep learning to revolutionize diverse application areas, ranging from traditional computer vision, and natural language processing to protein folding, and material discovery~\cite{kaddour2023challenges, lecun2015deep}. Only recently have large-scale optimizations become increasingly appreciated in the optics community, following the widespread adoption of photonic inverse design~\cite{molesky2018inverse}. An end-to-end formulation, whose backbone is gradient back-propagation, seamlessly merges disparate physical and inferential components, elevating photonic inverse design to the next level of computational scale and complexity, which could rival that of celebrated deep learning models. In fact, several tactics, which have been successfully applied to large language models (LLMs)~\cite{narayanan2021efficient}, could be adapted to \emph{hyper-scale} end-to-end differentiable ``physical-inferential'' models. These tactics would have to exploit a strategic mix of data, task, model, and pipeline parallelisms across multiple GPUs and CPUs, traversing complex computational graphs, which must efficiently manage nested optimization levels and embedded physics simulators in both forward and adjoint directions.

End-to-end inverse design also dovetails with the emerging trend of digital twins in many industrial and research enterprises. Metaoptics end-to-end models may be considered ``photonic digital twins,'' primarily employed during the design stage, but can also be integrated with interactive feedback during deployment to provide new levers of control that touches the most profound physical processes at the nanoscale. Such feedback can be realized by rapidly maturing phase change material and optoelectronics platforms, enabling even pixel-level control and reconfigurability~\cite{gu2024programmable}. From a broader perspective, end-to-end design and optimization could accelerate the advent of ultra-scale physical design models, which consist of fully differentiable, multi-scale, multi-physics partial and ordinary differential equation simulators nested with signal processing, inference, and control algorithms. Large end-to-end models could, in turn, disruptively innovate conventional photonics, electronics, optoelectronics, acoustics, and more, unlocking unprecedented levels of sensitivity, tunability, footprint, efficiency, and information capacity. At these extreme scales and complexity, we speculate ``emergent'' abilities leaping out of hyperscale end-to-end models, similar to what has been intriguingly observed in LLMs -- phenomena which have been likened to critical phase transitions in statistical physics~\cite{wei2022emergent}. 

\section{Conclusion}

In this Perspective, we have highlighted the transformative potential of integrating metaoptics with computational imaging. By combining the subwavelength, multifunctional control offered by metasurfaces with advanced computational techniques, we can overcome the limitations of traditional imaging systems and enable new functionalities. One of the central ideas of this Perspective is the end-to-end co-design of hardware and software, facilitated by gradient backpropagation and optimization methods -- which allows for the direct optimization of computational metaoptics performance metrics. Applications such as phase imaging and quantum photonic state measurement exemplify how this synergistic approach can lead to compact, efficient, and robust imaging solutions for complex imaging tasks. 

We believe that computational metaoptic imaging could have a profound impact across a range of fields. In biomedical imaging, the ability to perform advanced phase imaging in compact setups could advance diagnostics by enabling portable, point-of-care testing in resource-limited settings. In quantum technologies, computational metaoptics could enable efficient and scalable measurement of quantum states, facilitating the development of quantum communication and computation systems. In machine vision and autonomous systems, optical encoders designed through end-to-end optimization could enhance object recognition and scene understanding while reducing computational load in the digital domain.

Looking ahead, the most exciting and speculative directions involve the development of hyperscale end-to-end differentiable models -- which we anticipate will be enabled by the exponential growth in computational power and advancements in high-performance computing architectures, allowing for unprecedented large-scale optimization and simulation of complex physical systems. Such models may unlock emergent abilities and unprecedented levels of performance in imaging systems. Such advancements could lead to disruptive innovations not only in photonics but also in electronics, optoelectronics, and beyond. The convergence of metaoptics and computational imaging holds the promise of transformative breakthroughs that could reshape the landscape of imaging and sensing technologies.

\section{Competing interests}
The authors declare no potential competing financial interests.

\section{Data and code availability statement}
The data and codes that support the plots within this paper and other findings of this study are available from the corresponding authors upon reasonable request. Correspondence and requests for materials should be addressed to C.~R.-C. (chrc@stanford.edu) and Z.~L. (zinlin@vt.edu).

\section{Acknowledgements}
C.~R.-C. is supported by a Stanford Science Fellowship. 
K.~W. acknowledges the support from Qu\'ebec’s Minist\`ere de l’\'Economie, de l’Innovation et de l'Énergie (MEIE), Natural Sciences and Engineering Research Council of Canada (NSERC), [RGPIN-2023-03630], and Fonds de Recherche du Qu\'ebec - Nature et Technologies (FRQ-NT).

\bibliographystyle{ieeetr}
\bibliography{bibliography}

\end{document}